# Light-controllable hybrid aligning layer based on LIPSS on sapphire surface and PVCN-F film

I. Gvozdovskyy[a*], D. Bratova[b], Z. Kazantseva[c], S. Malyuta[b,c], P. Lytvyn[c], S. Schwarz[d], R. Hellmann[d]

[a]Institute of Physics, NAS of Ukraine, 46 Nauky ave., Kyiv, 03028, Ukraine

[b]National Technical University of Ukraine "Igor Sikorsky Kyiv Polytechnic Institute", 37 Peremohy ave., Kyiv, 03056

[c]V. E. Lashkaryov Institute of Semiconductor Physics, NAS of Ukraine, 41 Nauky ave., Kyiv, 03028, Ukraine

[d]University of Applied Sciences, Würzburger Straße 45, 63743 Aschaffenburg, Germany

E-mails:

igvozd@gmail.com (I. Gvozdovskyy); bratova.dr@gmail.com (D. Bratova); kazants@isp.kiev.ua (Z. Kazantseva); serhiy.malyuta@gmail.com (S. Malyuta); Plyt2007@gmail.com (P. Lytvyn); Simon.Schwarz@th-ab.de (S. Schwarz); Ralf.Hellmann@th-ab.de (R. Hellmann)

**ABSTRACT**

The creation of aligning layers for the uniform orientation of liquid crystals is significant for both research and the application of liquid crystals. For all applications, the creation of aligning layers possessing controllable characteristics such as azimuthal and polar anchoring energies, easy-axis of director alignment and pre-tilt angle, in the same way as it is achieved by using photoaligning layers processed by light, is very important. Here, aligning properties of hybrid aligning layers created on the basis of sapphire surfaces additionally coated by photoaligning layer of PVCN-F are studied. These hybrid layers possess the properties of the nano-structured sapphire layer and the photosensitive PVCN-F layer, and complement each other. The irradiation time dependence of the azimuthal anchoring energy of the hybrid layers is studied. By using certain experimental conditions during irradiation of hybrid layers, *e.g.,* polarisation of light and irradiation time, a minimum value of the azimuthal anchoring energy, close to zero, was obtained. Atomic force microscope studies of the irradiated hybrid layers were also carried out. It was found that the behavior of the contact angle of nematic droplets placed on treated sapphire surfaces are in good agreement with properties of hybrid aligning layers and parameters of structuring surface obtained from AFM images.

**1 Introduction**

The alignment of liquid crystals (LCs) plays a key role in their application and is a figure of merit in research and development. It is well-known that the alignment of LCs occurs owing to the interaction between molecules of the substrate surface and LCs [1-5]. To obtain homogeneous alignment of LCs, dedicated aligning layers should be used. However, the alignment of LCs can also occur without the use of any aligning layers [2,6].

There are many different techniques for the treatment of surfaces to obtain a homogeneous alignment of LCs [1-4], namely a rubbing process [1,5], photoaligning [2-4, 7-11], Langmuir-Blodgett layers [12,13], ion/plasma beam [14,15], and many other processes [16-24]. However, each of these techniques possesses advantages and drawbacks, as summarized in detail in Ref. [22]. Recently, a new method of LCs alignment by laser-induced period surface structures (LIPSS) on different materials has been reported on polymer [19,20], titanium [21,22], ITO [23], and sapphire [25].

It is also well known that the creation of anisotropic properties of layers, *e.g.,* creation of periodically nano-grooves [1,5], cross-linking or destruction of polymer owing to light [2-4] or forming of surface charges [1,2] on the surface by means of any technique of treatment is one of the main factors to obtain homogeneous alignment of LCs.

The main characteristics of aligning layers are, *e.g.*, the azimuthal and polar anchoring energies, easy axis of director and pretilt angle [1-4]. On the one hand, these characteristics depend on the materials used as aligning layers [1], and also on the interaction between the molecules of both aligning layer and LCs, thus ensuring homogeneous *planar* or *homeotropic* (vertical) alignment of LCs [1-4]. On the other hand, the value of these characteristics can be controlled by changing certain parameters of treatment (*e.g.,* pressure of roller, polarisation of light, angle of irradiation, wavelength of irradiation, laser fluence, scanning speed of the laser beam and *etc*).

The creation and study of various layers possessing controllable properties are an important step for different applications [13,26]. In that respect, most promising materials are photopolymers, possessing reversible photoreactions and forming so-called '*command*' aligning layers [26]. Recently, the photoalignment of LCs on surfaces of various chalcogenide glasses as promising materials was studied [27-29]. However, before the process of irradiation by light, the photoaligning layer was often characterized by inhomogeneous (random) alignment of LCs, contrary to homogeneous planar alignment obtained for instance by means of the rubbing technique. Of course, such inhomogeneous alignment of LCs is undesirable for applications. Obviously, the combination of alignment properties obtained by various techniques with the use of various materials offers the possibility to obtain new hybrid aligning layers (HALs) with properties that are more efficient than those of its antecedents.

Recently, the combination of various techniques for the aligning of LCs was applied and their alignment properties, for example in case of the rubbed photopolymer layer of the poly(vinyl cinnamate) with further irradiation by linearly-polarized UV light [30-32], the polymer and the pre-rubbed polymer treated by LIPSS [19,20] or various surfaces with nano-grooves obtained by LIPSS [21-23] further covered by polymer possessing strong azimuthal anchoring energy (AAE) without rubbing treatment [21,22,25], were studied. In contrast to our previous studies [21,22,25], we extend our research to study alignment characteristics by combined surfaces having both sapphire substrates treated by LIPSS and a photosensitive polymer in order to achieve the tuning of the main alignment parameters (*e.g.,* easy orientation axis, azimuthal anchoring energy) by light.

By taking into account the fact that for the rubbing technique several shortcomings exist, *e.g.*, the accumulation of both static charges and dust particles, the use of LIPSS that are also characterized by nano-grooves (similar to the rubbing technique) [21-23,25], may be regarded as alternative non-contact surface treatment similar to the photoalignment technique. Furthermore, similar to the photoalignment technique, various properties of aligning layers can be obtained by changing the parameters during treatment of surfaces by LIPSS [21,22], as studied in detail [21-23].

However, it should be noted that the nano-structuring of various surfaces (*e.g.,* metals or transparent dielectric materials) by LIPSS needs individual and detailed studies aimed at increasing the quality of nano-grooves being characterized by both certain depth A and period Λ (please note that the period is always limited to the choice of laser wavelength). In other words, technical parameters of LIPSS (*e.g,.* scanning speed and laser fluence) that are used for the processing of metals will be incorrect in case of structuring dielectric materials. To obtain a high-quality nano-structured surface based on LIPSS on transparent dielectric materials (*e.g.* sapphire, SiO$_2$), the process of the surface structuring is possible when periodic alternation of the nano-structured (1D-LSFL) and unstructured lines occur, looking like a "*zebra*" marked crosswalk. All of this can be probably referred to certain shortcomings of the LIPSS technique of LC alignment. The study of uniform orientation of the nematic E7 by *zebra*-like surfaces of sapphire substrates possessing various widths of unstructured gaps *L* was recently described in detail [25]. It was show that *zebra*-like surfaces can be successfully used to obtain the homogeneous alignment of LCs and characterised by AAE dependent on the width of the unstructured gap *L*.

Comparing non-contact techniques of alignment of LCs, it should be noted that for example of the sapphire substrate processed by LIPSS, the AAE value ($W_\varphi \sim 4.3 \times 10^{-6}$ J/m$^2$ [25]) is close to the AAE value of polyvinyl-4-fluorocinnamate (PVCN-F) and can be changed in the range of $W_\varphi \sim (4 - 15) \times 10^{-6}$ J/m$^2$ by UV exposure [33,34]. Obviously, due to approximately the same values of anchoring energy $W_\varphi$ obtained by photoalignment and LIPSS technique, the combination of aligning layers (so-called HALs) for orientation of LCs can be important from a scientific point of view and promising for various applications, such as optics and display technology *etc*.

The general idea of the studies described in this manuscript is shown in Figure 1. The use of sapphire substrates processed by LIPSS, by forming a *zebra*-like surface, and, for example, coated with PVCN-F layer (Figure 1 (a)), on the one hand, obviously will be characterized by homogeneous alignment, owing to the availability of nano-grooves on a *zebra*-like sapphire substrate, with some initial parameters (*e.g.,* AAE $W_\varphi$, direction of easy orientation axis). On the other hand, initial aligning parameters of the HAL may change and be controlled by parameters of the UV light exposure, *i.e.*, direction of the polarization and exposure time, owing to the use of photosensitive PVCN-F layer (Figure 1 (b)-(e)).

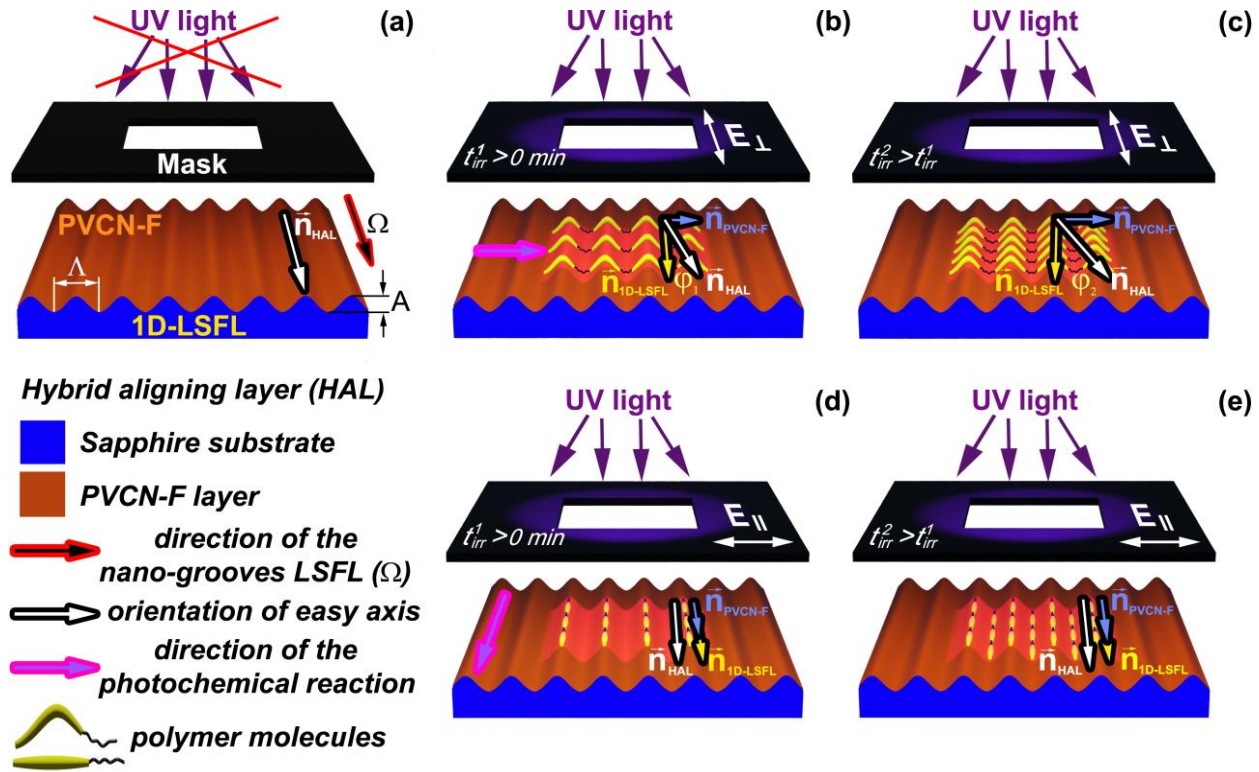

Fig. 1. Schematic of the studied substrates and their irradiation. (a) HAL, consisting of sapphire substrate (blue colour) coated by photosensitive PVCN-F layer (orange colour) before UV irradiation. The direction of the LSFL ($\Omega$) is shown by red and black arrow. $\vec{n}_{HAL}$ (black and white arrow) is the orientation of the easy axis of the HAL. HAL irradiated by polarized UV light with the polarization $E_\perp$ being parallel to the nano-grooves with their period $\Lambda$ and depth A over time $t_{irr}^1$ (b) and $t_{irr}^2 > t_{irr}^1$ (c). HAL irradiated by polarized UV light with the polarization $E_\parallel$ being perpendicular to the nano-grooves over time $t_{irr}^1$ (d) and $t_{irr}^2 > t_{irr}^1$ (e). $\vec{n}_{1D-LSFL}$ (black and yellow arrow) and $\vec{n}_{PVCN-F}$ (black and blue arrow) are directions of easy axis caused by structured sapphire substrate obtained by LIPSS and PVCN-F layer after UV-irradiation, respectively.

It is easy to assume that in HAL the presence of both nano-grooves of a *zebra*-like surface and photosensitive polymer layer will influence the orientation parameters of each other. The alignment of the HAL will take place due to the competition between anchoring caused by both the *zebra*-like surface and photosensitive polymer. It can be easily understood, for instance, by imagining a rower, who moves in windy conditions in such a way that the direction of the rower's motion is orthogonal to the direction of the wind. When the speed of the wind will exceed the speed of the rower, the preferred direction of the rower will be determined by the wind, and vice versa. In the case when the speeds are equal, no competition is observed. The preferred direction of the rower will be at 45 degrees. When the direction of rower and wind coincides then the motion speed of the rower will be additionally affected by the speed of the wind. In general, things like that will occur in case of the HAL. The competition between various alignments characterized by different aligning parameters could lead to various peculiarities in the orientation of LCs by such combined layers and will be described in this article.

## 2 Materials and methods

### 2.1 Materials

Nematic liquid crystal E7 obtained by Licrystal, Merck (Darmstadt, Germany) was chosen to study the aligning characteristics of the HAL. For the nematic E7, the optical and dielectrical anisotropy are $\Delta n = 0.2255$ ($n_e = 1.7472$, $n_o = 1.5217$) and $\Delta \varepsilon = +14.3$ ($\varepsilon_\parallel = 19.5$, $\varepsilon_\perp = 5.2$) at T = 20ºC, $\lambda$ = 589.3 nm, and f = 1 kHz. Furthermore, the elastic constants of the nematic E7 are $K_{11}$ = 11.7 pN, $K_{22}$ = 6.8 pN, $K_{33}$ = 17.8 pN for splaying, twisting and bending, respectively [35-38]. The temperature of the nematic-isotropic transition $T_{Iso}$ is 58 ºC [35].

To create the HAL both the structured transparent sapphire substrate [39] and the photosensitive polymer PVCN-F [40] were used.

To obtain planar alignment of the nematic liquid crystal E7, HAL based on a *zebra*-like surface of the structured sapphire substrate covered with polyvinyl-4-fluorocinnamate (PVCN-F) layer and glass substrates (microscope slides, made in Germany) covered with polyimide film PI2555 (HD MicroSystems, USA) were used.

### 2.2 Methods

The hybrid alignment layers are created on transparent dielectric substrates. Here, sapphire substrates (UQG Optics) having a thickness of 2 mm and an initial surface roughness of about 6 nm were used. To structure the surface of the substrates, a micromachining system (MM200-USP, Optec), being equipped with an ultrashort pulsed laser (Pharos, Light Conversion, Vilnius, Lithuania) was applied. As previously demonstrated [41], a diffractive beam shaper (ST-225-I-Y-A, Holo/Or) was used to transfer the Gaussian beam profile in a Top-Hat profile. Consequently, a more homogeneous intensity distribution and thus a higher quality of the generated laser-induced periodic surface structures LIPSS, *i.e.* low-spatial frequency LIPSS (LSFL) was realized in the focus plane. To move the laser across the samples, a 2D-galvano scanner in conjunction with a 100 mm f-θ-lens was applied. The Top-Hat had an edge length of 45 µm ($1/e^2$), while the laser parameters for the structuring process were the laser wavelength λ of 1030 nm, a repetition rate of 50 kHz and a pulse duration of 230 fs (FWHM).

To structure the surface of the sapphire substrates we used the experimental scheme described in [25]. For example, to cover an area of $5 \times 5$ mm$^2$ with LSFL, the laser scans a hatch consisting of parallel lines having width of 36 µm. The formed LSFL have a line width of 23 µm and therefore, an unstructured gap *L* between two LSFL lines has a width of 13 µm, making this hatch look like a crosswalk ("*zebra*"-like structured surface). In our studies, we used the structured sapphire surfaces having various hatch widths (30, 36 and 40 µm), resulting in different widths of the unstructured gap *L* (7, 13 and 17 µm).

As a photoaligning polymer, the well-studied polyvinyl-4-fluorocinnamate (PVCN-F) [33,34] was chosen. The synthesis of the PVCN-F was described in [40]. To cover glass or sapphire substrates by a polymer film, toluene solution of the PVCN-F (C = 10 g/L) was used.

To obtain the HAL, *zebra*-like surfaces of sapphire substrates were covered with PVCN-F by dipping technique. Therefore, the equipment for Langmuir-Blodgett film preparation R&K (Wiesbaden, Germany) was used. The sapphire substrates were dipped into the PVCN-F solution and subsequently lifted up vertically along the direction of the periodic nano-grooves of the LSFL. The speed for this process was set to 5 mm/min for a constant lifting. Furthermore, the HAL were annealed for 2 hours at a temperature of 80 °C in a Termolab SNOL 20/350 (Boryspil, Ukraine) laboratory dry oven to evaporate the solvent.

To obtain an aligning layer with strong AAE $W_\varphi$, the glass substrate was covered with a 10:1 n-methyl-2-pyrrolidone solution of the polyimide PI2555 by spin coating and further annealed as described in [42].

Figure 2 shows the typical setup of UV irradiation of sapphire substrates with HAL. As a source of UV radiation, the UV lamp DRT-240 (Kharkiv, Ukraine) emitting in a spectral range from 230 nm to 400 nm was used. The intensity of the UV radiation on the surface of the substrate was about 1.8 W/m$^2$. The distance between the UV lamp and the sapphire substrate was 50 cm. To obtain a high-quality alignment of the PVCN-F layer, polarized UV light was used. The linearly polarized UV irradiation was obtained using a Glan-Thompson prism with plane polarization $E_p$ in the plane of the substrate. In our studies, we used both vertical $E_\perp$ and horizontal $E_\parallel$ polarizations of UV light, which could be changed by rotating the Glan-Thompson prism by 90 degrees. The irradiation of the PVCN-F layer was carried out though the square-shaped Mask with a size of about $3 \times 3$ mm$^2$.

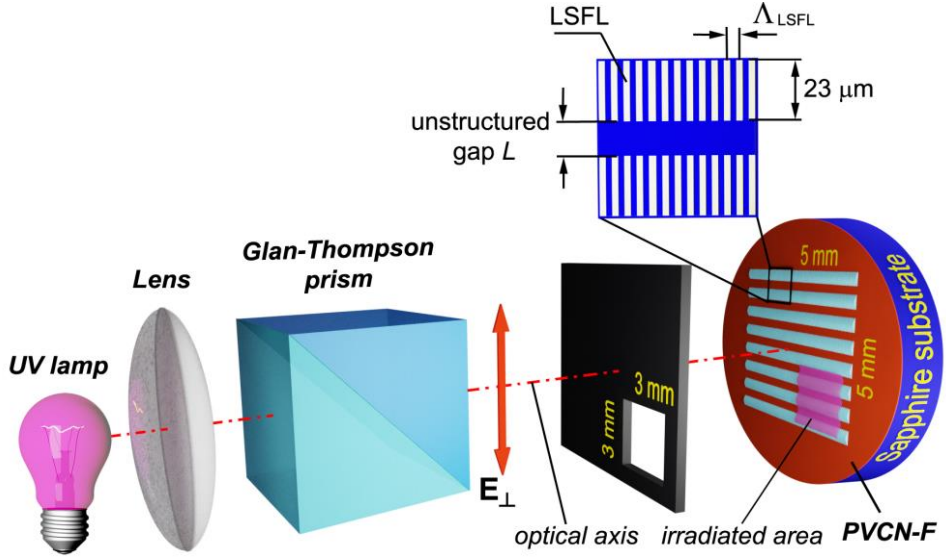

Fig. 2. Schematic setup of irradiation of the hybrid aligning layer based on the LIPSS on the sapphire and photoinduced aligning polymeric layer of the PVCN-F. The partial image of HAL schematically shows a *zebra*-like surface of sapphire substrate structured by LIPSS. A *zebra*-like surface is characterized by LSFL lines with 23 μm width, fixed period of nano-grooves $\Lambda_{LSFL}$ = 980 nm and different widths of unstructured gap $L$ (7, 13 and 17 μm). The intensity of the UV lamp was 1.8 W/m$^2$ on the substrate. The distance between the UV lamp and the substrate was 50 cm.

The well-known method of combined twist LC cells was used to measure the twist angle $\varphi$, as described in detail [22,25,42]. As a reference substrate we used the glass substrate covered with a PI2555 layer. To obtain the aligning layer characterized with strong AAE $W_\varphi$ (about $4 \times 10^{-4}$ J/m$^2$ [43]), the PI2555 layer was unidirectionally rubbed several times ($N_{rubb}$ = 15). The sapphire substrate with HAL was used as tested substrate of the combined twist LC cell.

To set a LC cell thickness range of 10 - 20 µm, we used Mylar spacers. The accurate measurement of thickness was carried out by the interference method, which was essentially about measuring the transmission spectrum of the empty cell with the spectrometer Ocean Optics USB4000 (Ocean Insight, USA, California).

To avoid possible flow alignment of the nematic E7, the LC cells were filled by capillary at elevated temperature of T = 61 °C, higher than the temperature of the isotropic phase transition [35] and slowly cooled to room temperature with a rate of about 0.1 °C/min.

The contact angle $\beta$ of the nematic E7 droplet placed on HAL was measured by the recently described method [44] being based on the determination of the linear dimensions of the liquid droplet at the surface with the use of a horizontal microscope. According to [45], the contact angle $\beta$ was calculated by using the measured diameter $D$ and height $H$ of the droplet being deposited on the HAL, as is follows:

$$\cos\beta = \frac{(D/2)^2 - H^2}{(D/2)^2 + H^2} \qquad (1)$$

A scanning electron microscope (SEM, Phenom ProX, PhenomWorld), an atomic force microscope (AFM, Dimension Icon, Bruker), and a transmitted light microscope (DM6000 M, Leica) were used to characterize the surface topography of the LSFL. The twist angles $\varphi$ of the combined twist LC cells were determined by polarizing optical microscope (POM) BioLar (PZO, Warsaw, Poland).

**3 Results and discussion**

*3.1 Azimuthal anchoring energy of the PVCN-F film*

In this section, the results of the twist angle $\varphi$ of combined twist LC cell consisting of the reference substrate with PI2555-coating and tested glass substrate covered with PVCN-F film having areas with various UV expositions are presented.

The photochemical reactions of PVCN-F films were studied detail in [40,46-48]. The scheme to irradiate the PVCN-F layer by UV light is shown in Figure 2. The plane of polarisation of the Glan-Thompson prism $E_p$ (in other words, horizontal $E_\parallel$ polarisation of UV light) was perpendicular to the long side of the tested substrate. It is well known that photochemical reactions (cross-linking of cinnamate groups) during irradiation of a PVCN-F film by linearly polarized UV light will occur in the direction perpendicular to the polarization of light due to the preferential orientation of the remaining cinnamate fragments [40,47], since the alignment of the LC molecules is perpendicular to the polarization of the UV light [40]. Each of the irradiated areas of the PVCN-F layer correspond to a certain irradiation time $t_{irr}$, which changes in the range from 0 to 30 min. Figure 3 shows the general view of the combined twist LC cell with tested substrate covered with PVCN-F layer possessing by areas (stripes) with various irradiation time $t_{irr}$.

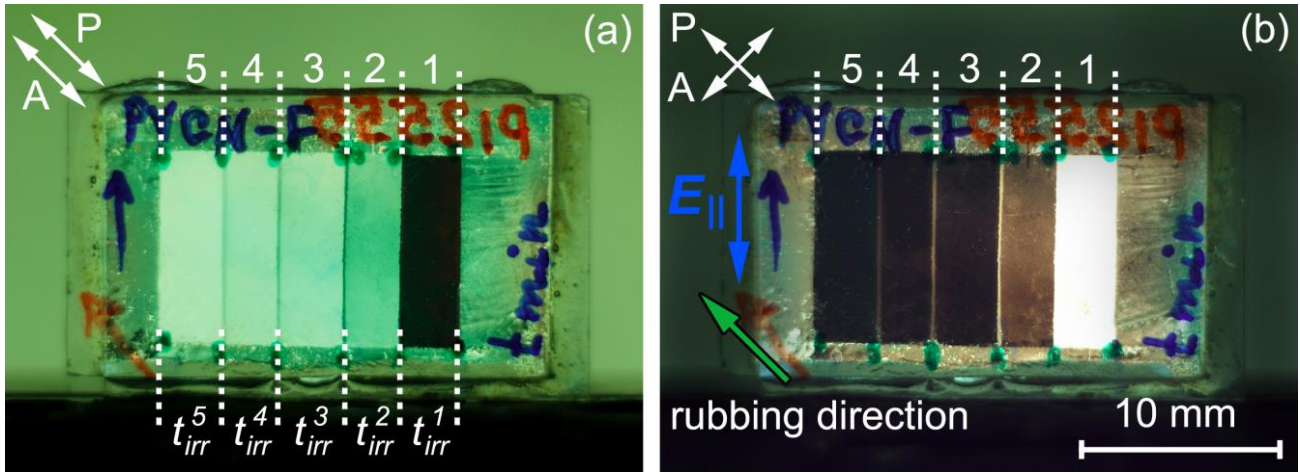

Fig. 3. Photographs of the combined twist LC cell, assembled from reference substrate with PI2555 layer, rubbed $N_{rubb} = 15$ at an angle of 45 degree, and tested substrate with PVCN-F coating between: (a) parallel and (b) crossed polarizer P and analyser A. Each area of PVCN-F layer was irradiated by: 1) $t_{irr}^1 = 1$ min; 2) $t_{irr}^2 = 3$ min; 3) $t_{irr}^3 = 6$ min; 4) $t_{irr}^4 = 10$ min; 5) $t_{irr}^5 = 30$ min. The intensity of the UV lamp was 1.8 W/m$^2$. The distance between the UV lamp and the tested substrate with PVCN-F layer was 50 cm. The plane of polarization of Glan-Thompson prism $E_\parallel$ is shown by blue arrow. The plane of polarization of the polarizer P coincides with the rubbing direction (black and green arrow) on the reference substrate. The thickness of the LC cell was 13.3 μm.

By rotating the analyser A at a certain angle $\varphi$ where the lowest transmittance of the light will be found, the value of this rotating angle is the twist angle of the LC cell. Each area of the combined twist LC cell, differed by UV irradiation time of the PVCN-F layer, has a certain value of the twist angle $\varphi$. The dependence of the twist angle $\varphi$ of the LC cell on the irradiation time $t_{irr}$ is shown in Figure 4 (a). Obviously, increasing the irradiation time $t_{irr}$ leads to the increase of the twist angle $\varphi$. However, in case of longer exposure times ($t_{irr}$ = 15 - 30 min) the twist angle $\varphi$ is almost unchanged.

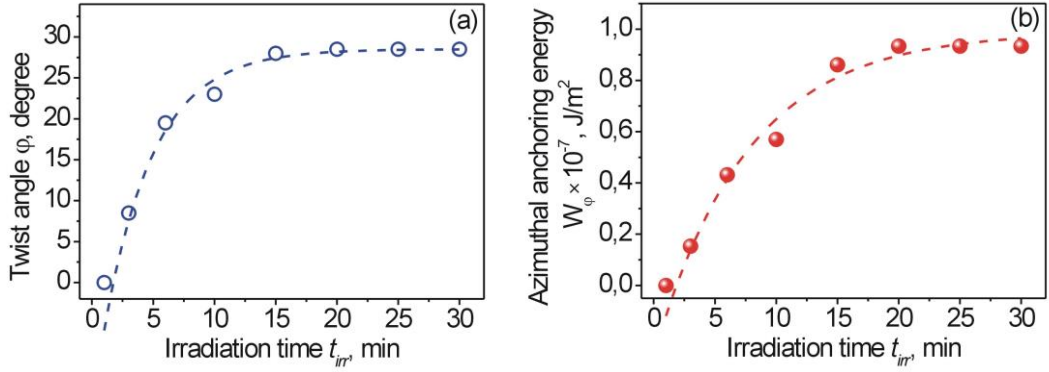

Fig. 4. Dependence of the twist angle $\varphi$ of the combined twist LC cell (a) and AAE $W_\varphi$ of the glass substrate covered with PVCN-F film (b) on the irradiation time $t_{irr}$. The intensity of the UV lamp was 1.8 W/m$^2$. The distance between the UV lamp and the tested substrate with PVCN-F layer was 50 cm. The thickness of the LC cell was about 13.3 μm. The dashed curve is a guide to the eye.

By knowing the measured value of the twist angle $\varphi$ of the LC cell, the value of AAE $W_\varphi$ of the irradiated PVCN-F layer is calculated by following Equation (2) [49,50]:

$$W_\varphi = K_{22} \times \frac{2 \times \varphi}{d \times \sin 2(\varphi_0 - \varphi)}, \qquad (2)$$

where $K_{22}$ is the twist elastic constant of the nematic E7, $d$ is the thickness of the LC cell, $\varphi_0 \approx 45°$ the angle between the easy axes of the reference and tested substrates and $\varphi$ is measured twist angle.

The dependence of AAE $W_\varphi$ of the PVCN-F layer on irradiation time is shown in Figure 4 (b). During longer UV exposure times, the value of AAE $W_\varphi$ reaches its maximum. This is due to the fact that a maximum quantity of molecules participates in the formation of long polymer chains (as schematically shown in Figure 1 (b) - (d)) owing to the cross-linking of cinnamate groups [40]. Thus, by obtaining the value of AAE $W_\varphi$ in the order of $1 \times 10^{-7}$ J/m$^2$, we can conclude that PVCN-F layer possessing a weak photoinduced anchoring [2].

*3.2 Azimuthal anchoring energy of the HAL*

By summarizing previous studies [21-23,25,42], it can be stated that the value of AAE $W_\varphi$ of the various aligning surfaces processed by LIPSS can change, on the one hand, by choosing of a certain type of aligning surface (*e.g.,* pure surface or surface covered with additional layers). On the other hand, the value of AAE $W_\varphi$ depends on the geometry of nano-structured surfaces such as the period $\Lambda$ and depth $A$ of nano-grooves [21-23,42] or, as in the case of transparent dielectric materials processed by LIPSS, the value of AAE $W_\varphi$ of a *zebra*-like surface also depends on the availability of certain periodic uniformities characterized by unstructured gaps between structured LSFL [25].

As calculated in [25] for a pure *zebra*-like surface of sapphire substrate (so-called first type of aligning layer or FTAL as set forth in Ref. [25]) and FTAL coated by ITO (so-called second type of aligning layer or STAL as set forth in [25]) the AAE $W_\varphi$ reaches a value in the order of $10^{-6}$ J/m$^2$. Therefore, the anchoring energy of a structured sapphire layer is about 10 times stronger than for PVCN-F layer (Figure 4 (b)). Due to the fact that for the STAL the planar alignment of LC is unstable (planar-homeotropic transition takes place [25]), this type of aligning layer will not be considered below.

For further studies we have chosen pure sapphire substrates characterized by *zebra*-like surfaces with different widths of the unstructured gaps $L$ (7, 13 and 17 µm) and, accordingly, values of AAE. Table 1 shows values of AAE $W_\varphi^\Sigma$ of *zebra*-like surfaces previously obtained in Ref. [25].

Table 1. Dependence of the AAE $W_\varphi^\Sigma$ of a *zebra*-like surface on width of the unstructured gap $L$. [25]

| Sample | Width of the unstructured gap $L$, µm | AAE $W_\varphi^\Sigma \times 10^{-7}$, J/m$^2$ |
|---|---|---|
| 1 | 7 | 9.7 |
| 2 | 13 | 7.7 |
| 3 | 17 | 5.3 |

As shown in Table 1, the AAE $W_\varphi^\Sigma$ of a *zebra*-like surface of the sapphire substrates are about 5 - 10 times higher than AAE ($W_\varphi \sim 0.86 - 0.93 \times 10^{-7}$ J/m$^2$) obtained for PVCN-F layers irradiated by UV light within 15 - 30 min (Figure 4(b)).

PVCN-F layer coating a *zebra*-like surface of the sapphire substrate forms the HAL, possessing properties of both the photosensitive PVCN-F film and the structured sapphire layer.

*3.2.1 AFM study of the HAL*

As previously shown in Ref. [25], the structured lines within the *zebra*-like surface of the structured sapphire substrates contain LSFL with a period $\Lambda$ of 980 nm and depth A of 100 nm. It was found in [25] that for the 1D-LSFL the value of AAE $W_\varphi^{struct}$ can be changed within the range of $(8.8 - 12) \times 10^{-7}$ J/m$^2$ depending on the width of the unstructured gap $L$. It should also be noted that the value of AAE $W_\varphi^{struct}$ for the 1D-LSFL, due to presence of nano-grooves, is always stronger than for the unstructured gap [25] or for the entire *zebra*-like surface, consisting of both 1D-LSFL and unstructured gaps (*e.g.* Table 1).

Now let us consider the change in geometrical dimensions of the nano-grooves for PVCN-F coated sapphire substrates. Figure 5 (a) shows the AFM image of the 1D-LSFL. Figure 5 (b) shows the cross-section of the 1D-LSFL taken from AFM for the HAL normalized to X-axis. In the 1D-LSFL areas, the HAL possesses periodic nano-grooves with the period $\Lambda$ of about 972 nm and depth A of about 50 nm. In the area of a *zebra*-like surface of sapphire substrate processed by LIPSS (the black square with size of $5 \times 5$ mm$^2$ in Figure 5 (c)), the quality of the homogeneous alignment of HAL is better than in the unstructured area characterized by random alignment of the LC (Figure 5(d)).

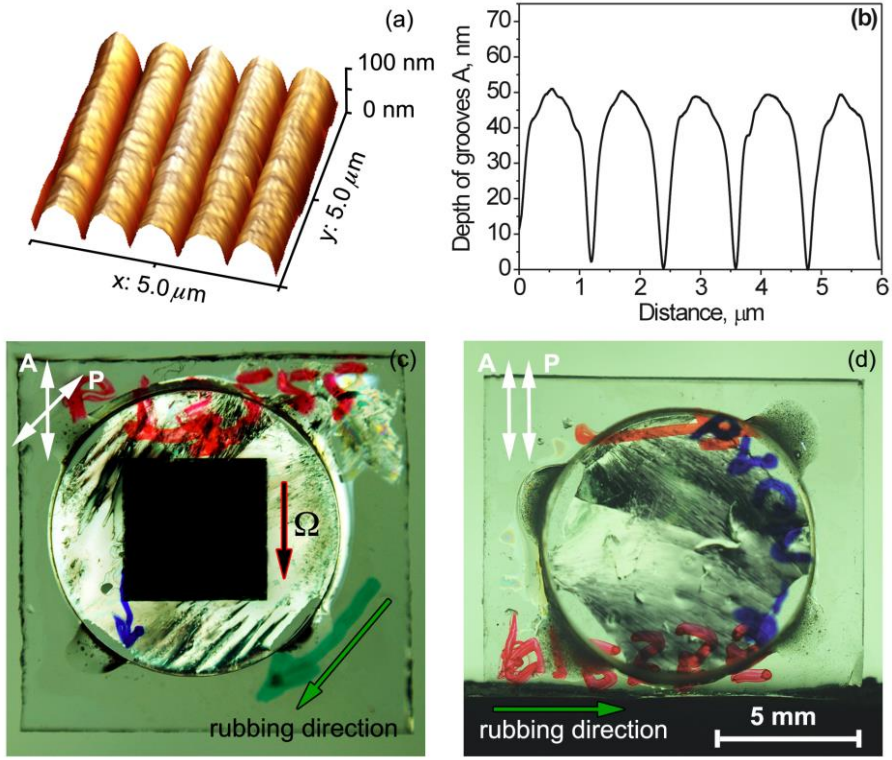

Fig. 5. (a) AFM image and (b) cross-section of the 1D-LSFL taken from AFM shown for the HAL, possessing periodic nano-grooves with period Λ ~ 972 nm and depth A ~ 50 nm. The cross-section is normalized to the X-axis. Photographs of combined twist LC cells, consisting of the reference unidirectionally rubbed glass substrate with PI2555 film coating and the tested substrates with: (c) a *zebra*-like surface coated with PVCN-F film, *i.e.,* HAL, and (d) the unstructured sapphire surface covered with PVCN-F before irradiation by UV light. Red and black arrow shows direction of nano-grooves of the LSFL $\Omega$. Black and green arrow shows the rubbing direction of the reference PI2555 film coating glass substrate. Thickness of LC cells was about 20 µm.

It was found that during UV irradiation of the HAL, depending on relative orientation between the polarization of UV light ($E_p$) and direction of the nano-grooves of 1D-LSFL ($\Omega$), the surface morphology varies in different ways (Figure 6). For instance, when the polarisation of the UV light $E_\perp$ is parallel to the $\Omega$ (Figure 1 (b), (c)), the decrease of the depth A of the nano-grooves occurs (Figure 6 (a), (b)) and vice versa when the polarisation $E_\parallel$ is perpendicular to the $\Omega$ (Figure 1 (d), (e)), the increase of the depth A takes place (Figure 6 (b), (c)). One of the main reasons of the variation of nano-grooves depth A is the change of the layer thickness, caused by photoinduced reactions in polymer films [51]. Depending on the polarization direction of linearly polarized UV light, the changes in physical properties such as transparency, dichroism and refractive index of the PVCN-F film were previously examined in [46-49]. It is assumed that the change of the surface morphology of the HAL in our studies is be caused by absorption dichroism of PVCN-F molecules [40,46,47] and specially given direction alignment of PVCN-F molecules along of the nano-grooves obtained by dipping technique (sapphire substrate with a *zebra*-like surface were dipped into the PVCN-F solution and lifted up vertically along the direction of the nano-grooves 1D-LFSL $\Omega$ [25]).

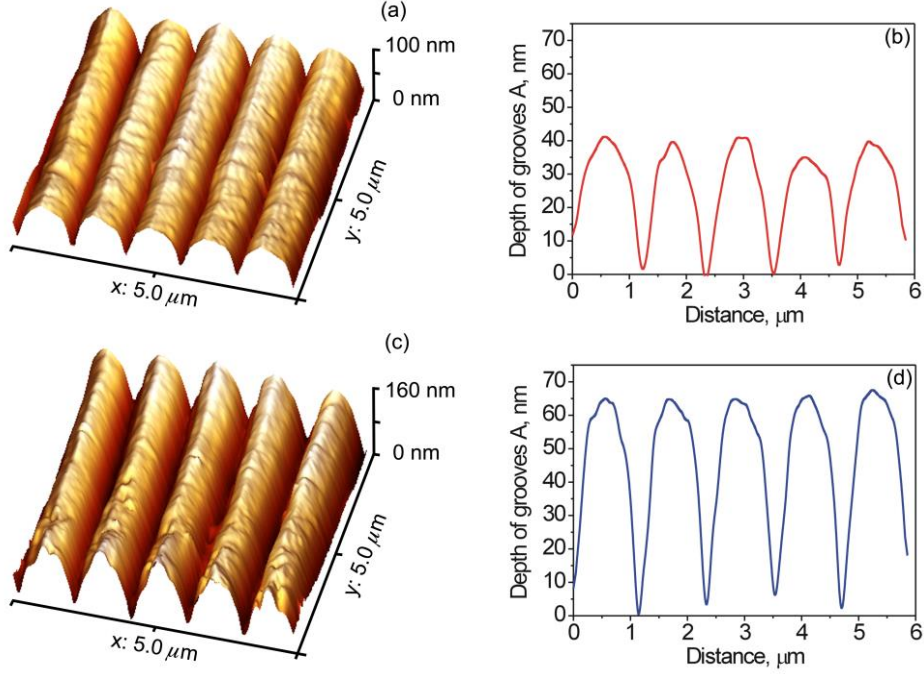

Fig. 6. AFM images and the cross-section of the 1D-LSFL taken from AFM reveal for the HAL irradiated by polarized UV-light with: (a), (b) the polarization $E_\perp$ parallel to the direction of the nano-grooves $\Omega$ of 1D-LSFL (period $\Lambda \sim 972$ nm and depth A $\sim 38$ nm) (c), (d) the polarization $E_\parallel$ perpendicular to the direction of the nano-grooves $\Omega$ of 1D-LSFL (period $\Lambda \sim 972$ nm and depth A $\sim 62$ nm). The irradiation time of the HAL was 15 min. The intensity of the UV lamp was 1.8 W/m². The distance between the UV lamp and the tested substrate with PVCN-F layer was 50 cm. The cross-sections of the areas with 1D-LSFL were normalized to the X-axis.

*3.2.2 Anchoring energy estimation of the HAL*

Let us qualitatively estimate the anchoring energy (AE) of a *zebra*-like surface of the structured sapphire surfaces by using the Berreman's theory [5,52] as shown in Equation (2) below:

$$W_B = 2\pi^3 \cdot K \cdot A^2 / \Lambda^3 \qquad (2)$$

where *K* is the arithmetical mean of the all Frank constants ($K_{11}$, $K_{22}$ and $K_{33}$) of the nematic LC, while A and $\Lambda$ are parameters (depth and period, respectively) of the nano-grooves of HAL obtained by AFM studies.

It should be noted that according to Equation (2) the AE $W_B$ of any aligning layers only depends on geometrical dimensions of nano-relief obtained under any processing of the surface of substrate (*e.g.* rubbing technique, photoalignment, LIPSS and others). The Berreman's theory does not take into account of the nature of interaction between the aligning surface and LC molecules leading to limitations in the accurate calculation of the anchoring energy. Since the irradiation of the HAL by polarized UV light with the polarization $E_\perp$ and $E_\parallel$ leads to various changes of geometrical dimensions of nano-relief (Figure 6), we have carried out an evaluation of the AE $W_B$ for these surfaces.

Recently, the qualitatively estimation of the value of AE $W_B$ for pure a zebra-like surface of sapphire substrate in both, the 1D-LSFL ($\sim 4.5 \times 10^{-6}$ J/m²) and unstructured gap ($\sim 0.97 \times 10^{-6}$ J/m²) was conducted in [25].

Table 2 shows the calculated AE $W_B$ of the HAL for both non-irradiated and irradiated surfaces by polarized UV light. By irradiation of the HAL the polarization of UV light was parallel or perpendicular with respect to the direction of the nano-grooves LSFL $\Omega$ (Figure 1(b)-(e)).

Table 2. AE $W_B$ of the HAL for non-irradiated surface and surfaces, irradiated by polarized UV light.

| Relative orientation of the UV light polarisation $E_p$ and direction of the nano-grooves $\Omega$ | Period of nano-grooves $\Lambda$, nm | Depth of nano-grooves A, nm | AE $W_B$, J/m² |
| --- | --- | --- | --- |
| Non-irradiated | 972 | 50 | $2.04 \times 10^{-6}$ |
| $E_\perp \parallel \Omega$ | 972 | 38 | $1.18 \times 10^{-6}$ |
| $E_\parallel \perp \Omega$ | 972 | 62 | $3.14 \times 10^{-6}$ |

As shown in Table 2 the UV irradiation of the non-irradiated HAL, possessing original depth A of nano-grooves ~ 50 nm, leads to a decrease or an increase of the depth A to values 38 nm and 62 nm, respectively. From AFM studies we can conclude that the value of depth A depends on the relative orientation of the UV light polarisation $E_p$ and direction of the nano-grooves of 1D-LSFL $\Omega$ (Table 2).

The minimum value AE $W_B$ ~ $1.18 \times 10^{-6}$ J/m² can be achieved through irradiation of HAL by polarized UV light when the polarization $E_\perp$ is parallel to $\Omega$. In this case, the photoinduced cross-linking of PVCN-F molecules occurs in a direction perpendicular to $\Omega$ (Figure 1(b), (c)). In comparison with the non-irradiated HAL (*i.e.,* depth A ~ 50 nm and $W_B$ ~ $2.04 \times 10^{-6}$ J/m²), the maximum value of AE $W_B$ ~ $3.14 \times 10^{-6}$ J/m² is achieved when the polarization $E_\parallel$ is perpendicular to $\Omega$. By taking into account the fact that the dichroism of absorption molecules [46,47], long axis of PVCN-F molecules oriented along the direction of the nano-grooves $\Omega$ and the polarization of UV light $E_\parallel$ is perpendicular to the $\Omega$ coincide (Figure 1(d),(e)), a significant photoinduced change of the surface morphology (*i.e.* depth A) is observed. In each of these cases of the irradiation of HAL by polarized UV light within the same irradiation time $t_{irr}$, the nano-grooves depth A was changed by an identical amount $\Delta A \approx A^{non-irr} - A_{t_{irr}}^{E_\perp \parallel \Omega} \approx A_{t_{irr}}^{E_\parallel \perp \Omega} - A^{non-irr}$ as can be easily estimated from Figure 5(b) and Figure 6(b),(d).

*3.2.3 Characterization of the HAL irradiated by UV light with the polarization $E_\perp$ parallel to the direction of the nano-grooves of 1D-LSFL $\Omega$*

In this subsection we present data of the measured twist angle $\varphi$ and calculated AAE $W_\varphi$ of the combined twist LC cells, consisting of the reference substrate with rubbed PI2555 layer and tested substrate with HAL, which was irradiated by polarized UV light with the polarization $E_\perp$ within various time.

To prepare the HAL, we used the nanostructured sapphire substrates, possessing various unstructured gap widths $L$ (7, 13 and 17 μm) but the same width of 1D-LSFL (Figure 2). The irradiation of the HAL is carried out with the use of the scheme of Figure 2, where the polarization of UV light $E_\perp$ is parallel to the direction of the nano-grooves of 1D-LSFL $\Omega$.

Total values of the twist angle $\varphi$ and the AAE $W_\varphi$ of the HAL consist of two components, namely, the nanostructures of the sapphire substrate and the PVCN-F layer irradiated by polarized UV light.

Let us consider the HAL based on the nanostructured sapphire substrate, having the width of the unstructured gap $L = 7$ µm and being covered with a PVCN-F layer. The dependence of the twist angle $\varphi$ of the LC cell on the irradiation time $t_{irr}$ is shown in Figure 8 (a). The total value of the twist angle $\varphi$ of the LC cell in the non-irradiated area of the HAL (opened black circles, line 1) is approximately 31.5 degrees. UV irradiation of the HAL leads to the change of the total twist angle $\varphi$ of the LC cell as shown in Figure 8 (a) by red spheres (*i.e.* curve 2). At the beginning of the UV irradiation of HAL with short expositions (*e.g.* $t_{irr} = 1 - 2$ min), a small change of the twist angle $\varphi$ is observed. During long expositions ($t_{irr} > 3$ min), the twist angle $\varphi$ of the combined twist LC cell (red spheres, curve 2) is decreased. Obviously, the decrease of the twist angle $\varphi$ occurs due to the photoinduced reaction of the PVCN-F molecules. However, as shown in Figure 4 (a) for a glass substrate covered with PVCN-F layer, the increase of the UV irradiation time leads to the increase of the twist angle $\varphi$. In the case of HAL, due to the certain choice UV irradiation conditions (*i.e.* when $E_\perp \parallel \Omega$), the direction of photoinduced cross-linking of PVCN-F molecules is occurring perpendicular to both the polarization of UV light $E_\perp$ [40,47] and direction of the nano-grooves of 1D-LSFL $\Omega$. It can be assumed that for substrates with HAL, the alignment of the LC molecules ($\vec{n}_{HAL}$) is the result of the competition between two alignments caused by a *zebra*-like surface of the sapphire substrate ($\vec{n}_{1D-LSFL}$) and the photoinduced cross-linking of the PVCN-F molecules ($\vec{n}_{PVCN-F}$), which have orthogonal directions to each other (Figure 1 (b), (c)).

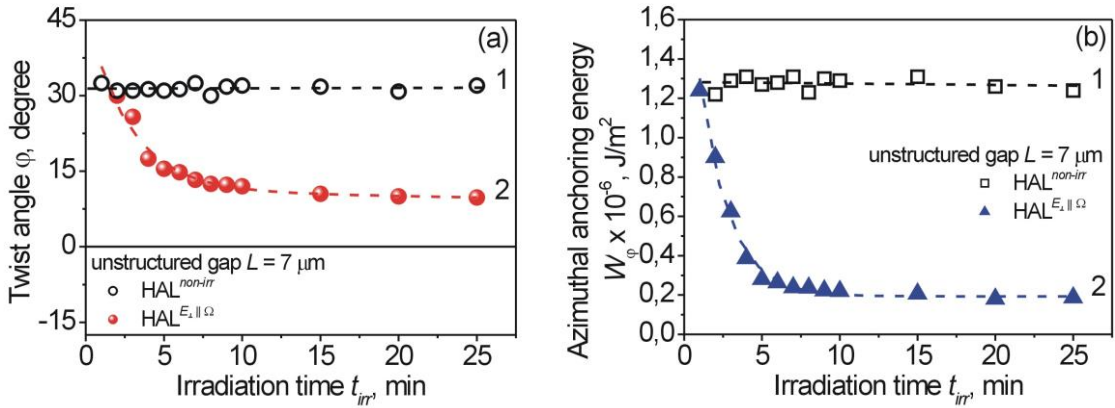

Fig. 7. The irradiation time dependence of the twist angle $\varphi$ (a) and AAE $W_\varphi$ (b) of the combined twist LC cell, consisting of the tested substrate, having both non-irradiated HAL (opened black symbols, lines 1) and HAL (solid symbols, curves 2), irradiated by polarized UV light with the polarization $E_\perp$ parallel to the $\Omega$. HAL with a *zebra*-like surface of the structured sapphire substrate has 7 μm width of the unstructured gap. Each point corresponds to a single LC cell. The average thickness of all LC cells was about 15.5 µm. The intensity of the UV lamp was 1.8 W/m$^2$. The distance between the UV lamp and the tested substrate with HAL was 50 cm. The dashed lines and curves are a guide to the eye.

It should be noted that in the case of the combined twist LC cell, the rubbing direction of the reference PI2555 substrate was carried out at an angle of 45 degrees to the direction of both, the nano-grooves of 1D-LSFL $\Omega$ and the photoinduced cross-linking of cinnamate groups [40] of the PVCN-F molecules.

From Figure 7 (a) we can conclude that for the HAL, having a width of the unstructured gap $L = 7$ µm, the non-zero total twist angle $\varphi$ is always observed. It is caused by stronger anchoring between LC molecules and HAL, induced by nano-grooves than photoinduced anchoring between LC molecules and PVCN-F layer.

Obviously, the difference between the twist angle $\varphi$ of non-irradiated HAL (Figure 7 (a), opened black circles) and the twist angle $\varphi$ obtained for the HAL irradiated by UV light (Figure 7 (a), red spheres), corresponds to the value of twist angle $\varphi$ of PVCN-F layer, irradiated by polarized UV light.

Figure 7 (b) shows the dependence of the AAE $W_\varphi$ of the non-irradiated (opened black squares, line 1) and irradiated HAL (solid blue triangle, curve 2) on irradiation time $t_{irr}$. For the irradiated HAL, the increase of irradiation time $t_{irr}$ leads to the decrease of the value of AAE $W_\varphi$. This behaviour is caused by photoinduced increase in the anchoring energy of the PVCN-F layer.

Figure 8 shows the comparison of the values of the twist angles between on the one hand, the difference obtained between the value of twist angles for the non-irradiated (curve 1, Figure 7(a)) and irradiated (curve 2, Figure 7(a)) areas of the HAL $\phi = \varphi_{non-irr}^{HAL} - \varphi_{irr}^{HAL}$ and, on the other hand, the value of the twist angle $\varphi^{PVCN-F}$ of the PVCN-F layer coating onto a glass substrate (Figure 4). The inset shows the difference $\Delta\varphi$ between the values of the twist angle $\phi$ (*i.e.* the twist angle caused only by photoinduced change of anchoring energy of the PVCN-F layer as a part of the HAL) and twist angle $\varphi^{PVCN-F}$ for a glass substrate covered with PVCN-F layer (*i.e.* $\Delta\varphi = \varphi^{PVCN-F} - \phi$). One of the reasons for this difference $\Delta\varphi$ may be the use of LC cells with different thicknesses.

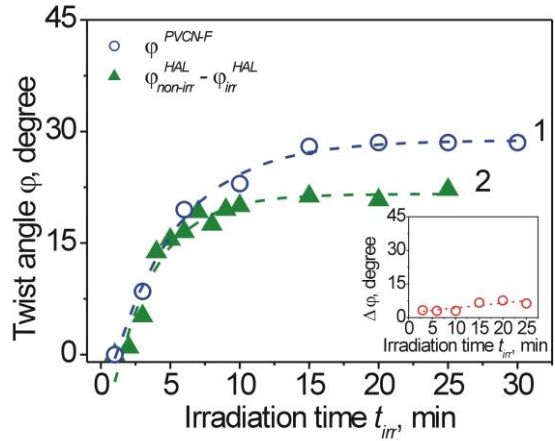

Fig. 8. The dependence of the twist angle $\varphi^{PVCN-F}$ of the combined twist LC cell, consisting of the tested PVCN-F layer coated onto a glass substrate (opened blue circles, curve 1) and the difference $\phi$ between the twist angle values of the combined twist LC cell consisting of the tested substrate with HAL having non-irradiated and irradiated areas (solid green triangles, curve 2), on the irradiation time $t_{irr}$. The inset shows $\Delta\varphi(t_{irr})$. The dashed curves are a guide to the eye.

Obviously, under these irradiating conditions (*i.e.* when $E_\perp \parallel \Omega$), the usage of a *zebra*-like surface of sapphire substrates having smaller AAE $W_\varphi$ will allow HAL that will be characterized by near zero twist angle $\varphi$ and as a result weak AAE $W_\varphi$.

Figure 9 shows the dependence of the twist angle $\varphi$ of the twist LC cell on the irradiation time $t_{irr}$, while the tested substrate HAL had a *zebra*-like surface with an unstructured gap $L = 13$ µm and 17 µm.

For non-irradiated areas of the HAL (*i.e.* unstructured gaps $L = 13$ and 17 µm), twist angles $\varphi$ of LC cells are unchanged for each gap width (opened black circles, curves 1, Figure 9). However, for various unstructured gaps the twist angles are different (*i.e.* $\varphi \sim 31.5°$ for $L = 13$ µm and $\varphi \sim 27.6°$ for $L = 17$ µm) due to different width of the gap $L$. This observation is in good agreement with the results obtained in [25].

In addition, twist angle values can go through zero during irradiation of the HAL with polarized UV light (Figure 9, blue and green spheres, curves 2). Therefore, by extrapolating data of fitting exponential curves 2, we can see that the near zero value of the twist angle $\varphi$ of LC cells is observed after irradiation of the HAL for about 12 min. Figure 9 shows the experimentally obtained values of the twist angles $\varphi$ of combined twist LC cells that are close to zero by solid red stars (*i.e.* ~ 2° for $L = 13$ µm and ~ 1.2° for $L = 17$ µm).

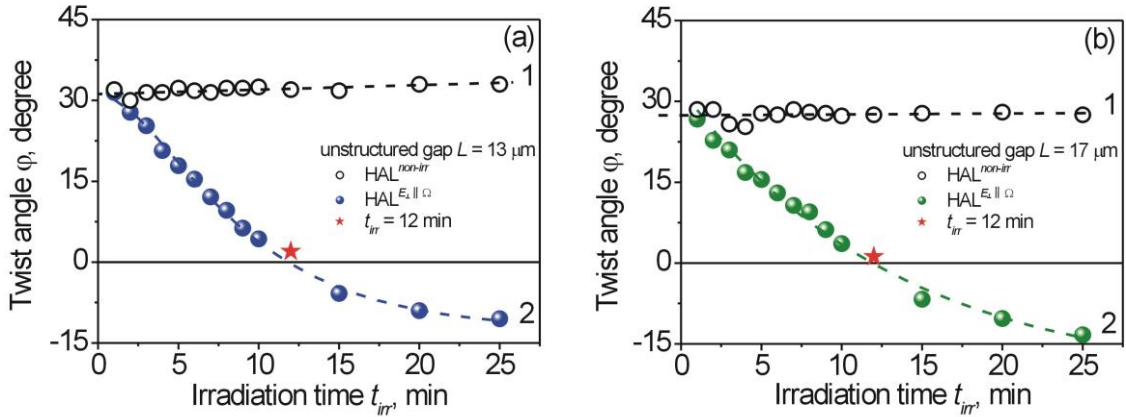

Fig. 9. The irradiation time dependence of the twist angle $\varphi$ of the combined twist LC cells consisting of the tested sapphire substrate with HAL. The twist angle of LC cells for non-irradiated area of the HAL (opened black symbols, lines 1) is: (a) $\varphi \sim 31.5°$ for $L = 13$ µm and (b) $\varphi \sim 27.5°$ for $L = 17$ µm. The twist angle of the LC cells with the HAL irradiated with polarized UV light when the polarization $E_\perp$ is parallel to the $\Omega$ (spheres, curves 2). Solid red stars correspond to twist LC cells having substrate with HAL irradiated by UV light for 12 min. Each point within the plots corresponds to each single LC cell. The average thickness of all LC cells was about 15.4 µm. The intensity of the UV lamp was 1.8 W/m$^2$. The distance between the UV lamp and the tested substrate with HAL was 50 cm. The dashed lines and curves are a guide to the eye.

Non-zero values of the twist angle $\varphi$ (*i.e.* 2 and 1.2 degrees) can be explained by inaccuracy of the polarization direction of UV light during irradiation of the HAL and further preparation of combined twist LC cell, and eventually error of measurement of the twist angle $\varphi$ by using the scheme described in Ref. [22].

Figure 10 shows the experimentally obtained dependence $W_\varphi(t_{irr})$ for the combined twist LC cells assembled with tested substrates with HALs having unstructured gaps of $L = 13$ µm width (solid red triangles, curves 2) and $L = 17$ µm (solid blue triangles, curves 2), which were irradiated by polarized UV light.

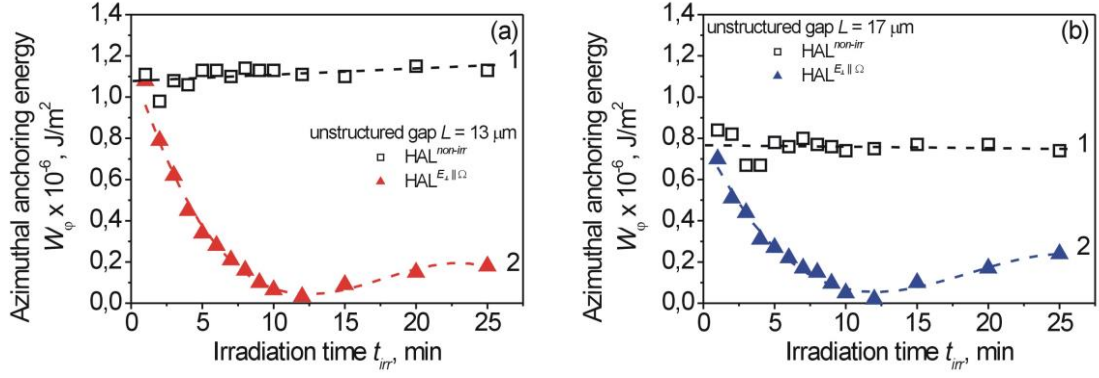

Fig. 10. Dependence of the AAE $W_\varphi$ of the combined twist LC cells on irradiation time $t_{irr}$. LC cells consist of the tested substrate with HAL, possessing structured sapphire surface with the width of the unstructured gap: (a) $L$ = 13 µm and (b) $L$ = 17 µm. The AAE $W_\varphi$ of the non-irradiated HAL (opened black squares, lines 1) is: (a) ~ $1.1 \times 10^{-6}$ J/m$^2$ for $L$ = 13 µm and (b) ~ $0.76 \times 10^{-6}$ J/m$^2$ for $L$ = 17 µm. The minimum of the value of AAE $W_\varphi$ for the HAL irradiated within 12 min by UV light with the polarization $E_\perp$ parallel to the direction of LSFL $\Omega$ (solid triangles, curves 2) is: (a) ~ $3.2 \times 10^{-8}$ J/m$^2$ for $L$ = 13 µm and (b) ~ $2 \times 10^{-8}$ J/m$^2$ for $L$ = 17 µm. Each point of the dependence corresponds to each single LC cell. The average thickness of all LC cells was about 15.4 µm. The dashed lines and curves are a guide to the eye.

In the case of the non-irradiated HAL (opened black squares, lines 1), the value AAE $W_\varphi$ is about 1.1 and 0.76 $\times 10^{-6}$ J/m$^2$ for width of the unstructured gap $L$ = 13 and 17 µm, respectively. The difference between the experimentally obtained values of AAE $W_\varphi$ (Figure 10) and theoretically calculated by Equation (2) AE $W_B$ (Table 2) for the substrate with non-irradiated HAL can be assigned to the fact that, on the one hand, the Berreman's theory only includes the geometrical dimensions of the nano-relief, such as period $\Lambda$ and depth A, of the nano-grooves. On the other hand, this theory does not consider the anchoring caused by physical and chemical interaction between the aligning layer (*e.g.* pure surface or surface covered with polymer layers) and the LC molecules.

AAE $W_\varphi$ values of the non-irradiated surface with HAL are about 3 times less than for the case of a *zebra*-like surface sequentially covered with both ITO and PI2555 layers (so-called third type of aligning layer or TTAL) recently described in Ref. [25]. The main reason of this difference is the fact that *zebra*-like surfaces of the sapphire substrates were covered with various polymers, which in turn have various values of AAE $W_\varphi$ (again we recall that PI2555 layer has stronger anchoring [42] than PVCN-F-layer [34]).

Figure 10 shows that an increasing exposure time ($t_{irr}$) during the UV irradiation of the HAL leads to a decrease of both, the twist angle $\varphi$ and the value of the AAE $W_\varphi$. The minimum values of AAE $W_\varphi$ are 3.2 and $2 \times 10^{-8}$ J/m$^2$, which are achieved under 12 min exposure of HALs. During 12 min UV irradiation, the photoinduced anchoring of the PVCN-F layer and the anchoring caused by a *zebra*-like surface are close to each other (*i.e.* it is similar to what happens in the case when the speeds of the rower and the wind are equal). However, the direction of the photoinduced easy axis ($\vec{n}_{PVCN-F}$) and the direction of the easy axis caused by nano-grooves ($\vec{n}_{1D-LFSL}$) are mutually orthogonal (Figure 1 (b),(c)). For these irradiation conditions (*i.e.* when the polarization of UV light $E_\perp$ is parallel to the $\Omega$) HALs characterized by the width of an unstructured gap $L$ = 13 and 17 µm have weak AAE $W_\varphi$ and consequently, combined twist LC cells very weakly rotate the input polarized light (*e.g.* by 2 and 1.2 degrees, respectively). It can be assumed that due to the long-range orientational order of LC molecules, the direction of the easy axis ($\vec{n}_{HAL}$) at the irradiated area of HAL is almost the same as for the rubbed PI2555 layer of the reference substrate.

Figure 11 shows the combined twist LC cell possessing a width of 17 µm of the unstructured gap and irradiated by polarized UV light (*i.e.* when the polarization $E_\perp$ is parallel to the $\Omega$) for 12 min.

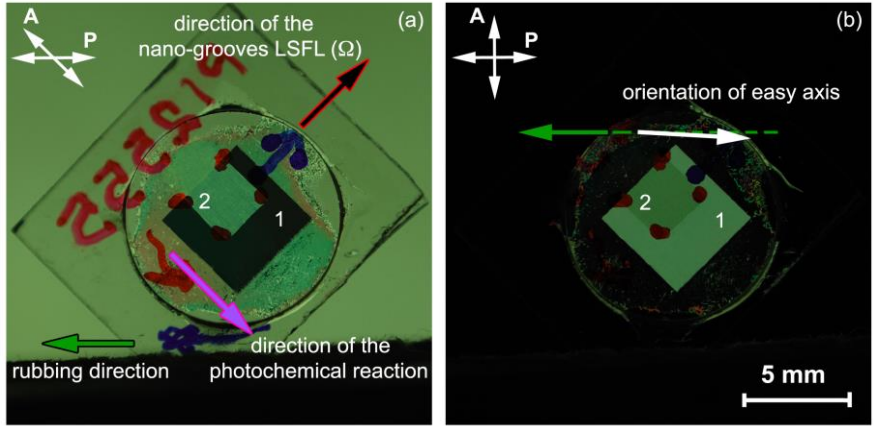

Fig. 11. Photograph of the combined twist LC cell, consisting of the reference substrate with rubbed PI2555 layer ($N_{rubb}$ = 15) and tested sapphire substrate with HAL, possessing the 17 µm width of unstructured gap, placed between crossed polarizer (P) and analyzer (A) at angles: (a) 27.5 and (b) 90 degrees. HAL possesses two areas: (1) non-irradiated and (2) irradiated by linearly polarized UV light (*i.e.* when the polarization $E_\perp$ is parallel to the $\Omega$) for 12 min. The thickness of the LC cell was 15.5 µm.

In the photographs, both, non-irradiated (area 1) and irradiated (area 2) HAL are shown. As described in [25], the orientation of easy axis of a *zebra*-like surface of a structured sapphire substrate is almost the same as the direction of the nano-grooves LSFL $\Omega$ (red and black arrow, Figure 11 (a)). In this case the deviation of the easy axis of LC alignment from the nano-grooves direction depends on the AAE $W_\varphi$ of the HAL and is approximately 17.5 degrees. The twist angle $\varphi$ of the 15.5 µm LC cell is about 27.5 degrees (Figure 9 (b), line 1). For the LC cell placed between P and A, crossed at an angle of -27.5 degrees, the minimum transmission of light is observed in the area 1 characterized by the non-irradiated HAL (Figure 11 (a)). When the angle between the polarizer P and analyzer A is 90 degrees, the area 1 transmits linearly polarized light owing to the rotation of the polarization plane (so-called Maugen regime [53]) which is caused by various alignment of the LC molecules on both, the reference (rubbed PI2555 layer) and the tested (*i.e.* surface with HAL) substrates (Figure 11 (b)).

Therefore, the twist angle $\varphi$ (*i.e.* orientation of easy axis) or AAE $W_\varphi$ can be changed by the geometrical parameters of nanostructured surfaces (*i.e.* period $\Lambda$, depth A and width of the unstructured gap $L$) [21-25] and the UV irradiation of the HAL with the photoinduced aligning layers (*e.g.* for PVCN-F layers curves 2, Figure 9).

In the irradiated HAL (area 2, Figure 11), the easy axis of the LC alignment is characterized by small twist angle $\varphi$ = 1.2 degrees (Figure 9 (b), red star). In this case, due to weak anchoring of HAL, being irradiated for 12 min, the greatest deviation of the easy axis of LC alignment from the direction of the photoinduced cross-linking of PVCN-F molecules (red and violet arrow, Figure 11 (a)) is about 43.5 degrees. This deviation angle is almost the same as the angle between the rubbing direction (green arrow, Figure 11 (a)) of the reference substrate (rubbed PI2555 layer) and the direction of the photochemical reaction of the PVCN-F molecules. Thus, due to the weak AAE $W_\varphi$ of the HAL in the area 2 of the LC cell, a small twist angle $\varphi$ = 1.2 degrees is observed. Figure 11 (b) shows by the white arrow the direction of the LC alignment in the area 2. In this area, a weaker transmission of light compared with area 1 is observed (Figure 11 (b)).

*3.2.4 Characterization of the HAL irradiated by polarized UV light with the polarization $E_\parallel$ perpendicular to the direction of the nano-grooves of 1D-LSFL $\Omega$*

In this subsection, we describe the changes of both, the twist angle $\varphi$ of LC cell and the value of the AAE $W_\varphi$ of HAL being irradiated by UV light with the polarization $E_\parallel$ perpendicular to the $\Omega$ depending on the irradiation time $t_{irr}$.

As mentioned above, the photoinduced cross-linking of PVCN-F molecules is in the direction perpendicular to the polarization of UV light [40,46-48]. For current irradiation conditions, the phototransformation of PVCN-F molecules will happen in the direction of the nano-grooves of 1D-LSFL $\Omega$ (Figure 1 (d),(e)), which causes the growing of the depth A (Figure 6 (b)).

Figure 12 shows the dependence of the twist angles $\varphi$ of the combined twist LC cells on the irradiation time $t_{irr}$ for various areas of the HAL. The measurements of the twist angles $\varphi$ were carried out using POM for irradiated (Figure 12 (a)) and non-irradiated (Figure 12 (b)) areas of the HAL, including 23 µm width structured 1D-LSFL (solid blue symbols, curves 1) and 17 µm width unstructured gap (opened green symbols, curves 2). The values of the twist angles of the structured 1D-LSFL ($\varphi_{irr}^{str}$, curve 1) and the unstructured gap ($\varphi_{irr}^{unstr}$, curve 2) are different, being in agreement with obtained results in [25]. In addition, the twist angle of the HAL (*i.e.* a *zebra*-like surface of sapphire substrate covered with PVCN-F layer) is the average value between the twist angles of structured 1D-LSFL and the unstructured gap, namely $\langle \varphi_{irr}^{HAL} \rangle = (\varphi_{irr}^{str} + \varphi_{irr}^{unstr})/2$ (spheres, curve 3, Figure 12).

During irradiation of the HAL with polarized UV light for short exposure times $t_{irr}$ (shorter than 7 min), the monotonous increase of the twist angle of both the structured 1D-LSFL (solid blue circles, curve 1) and the unstructured gap (opened green circles, curve 1) is observed. In case of the long-term irradiation ($t_{irr} > 7$ min) by polarized UV light, the value of the twist angle $\varphi$ for all areas is not changed. The main reason for this behaviour is the appearance of the maximum quantity of photoinduced cross-linking of PVCN-F molecules in the HAL [39,46].

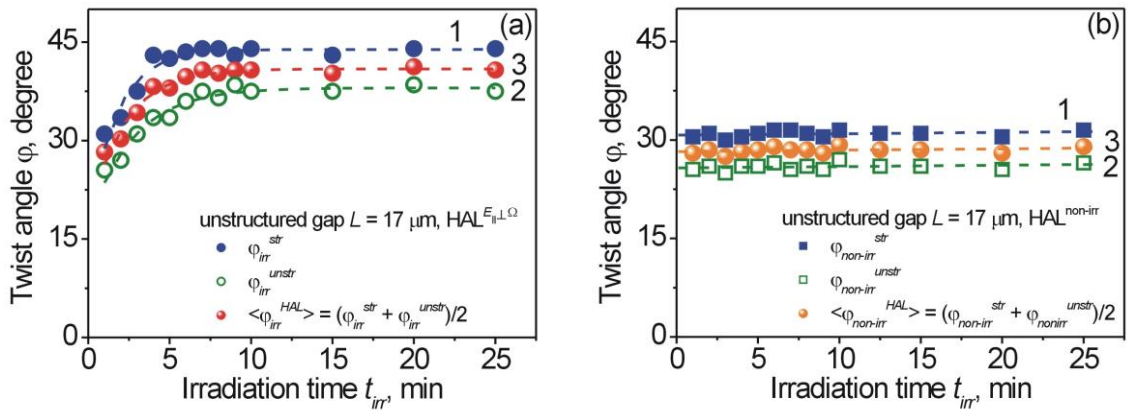

Fig. 12. Dependence of the twist angles $\varphi$ on the irradiation time $t_{irr}$ for the combined twist LC cells consisting of the reference substrate (rubbed PI2555 layer) and the tested sapphire substrate with: (a) irradiated by UV light with the polarization $E_\parallel$ perpendicular to the $\Omega$ and (b) non-irradiated areas of the HAL. The change of twist angle for: 23

µm width nano-grooves 1D-LSFL (solid blue symbols, curves 1), 17 µm width unstructured gap (opened green symbols, curves 2) and their average value (spheres symbols, curves 3). HAL consists of a *zebra*-like surface of the sapphire substrate (*i.e.* 23 µm 1D-LSFL and unstructured gap $L$ = 17 µm) covered with the PVCN-F layer. Each point of the dependence corresponds to each single LC cell. The average thickness of all LC cells was about 14.8 µm. The intensity of the UV lamp was 1.8 W/m$^2$. The distance between the UV lamp and the tested substrate with HAL was 50 cm. The dashed curves and lines and curves are a guide to the eye.

Figure 12 (b) shows the measured twist angles in the non-irradiated area of the HAL for various LC cells, which are characterized by different exposure times $t_{irr}$ in the irradiated area of the HAL with the size of $3 \times 3$ mm$^2$. It is obvious that in this case the twist angles $\varphi$ should not change. In fact, Figure 12 (b) shows the deviation of the twist angle $\varphi$ in the non-irradiated area for various LC cells, and these data demonstrate the accuracy of measurements and thoroughness of preparation of the LC samples. As described in Ref. [25] owing to the availability of nano-grooves of structured 1D-LSFL, the twist angle $\varphi_{non-irr}^{str}$ (solid blue squares, curve 1) is larger than the twist angle $\varphi_{non-irr}^{unstr}$ for the unstructured gap (opened green squares, curves 2). In addition, the calculated average value of the twist angle $\langle \varphi_{non-irr}^{HAL} \rangle$ = 28.4 degrees (orange spheres, curve 3) is almost the same as the twist angle ($\varphi$ = 27.6 degrees) measured for a *zebra*-like surface of the non-irradiated HAL (opened black squares, line1 in Figure 9 (b)).

Figure 13 shows the dependence of $W_\varphi(t_{irr})$ for the different areas of the HAL, namely non-irradiated (opened blue diamonds, line 1) and irradiated by polarized UV light (solid red diamonds, curve 2). The calculation of the AAE $W_\varphi$ was carried out for both, the non-irradiated and the irradiated areas of the HAL for a *zebra*-like surface of structured sapphire.

Obviously, in the non-irradiated area of HAL, no change of the AAE $W_\varphi$ is observed (line 1, Figure 13). However, during irradiation of the HAL the monotonous growth of the AAE $W_\varphi$ with increasing of the exposure time is observed (curve 2, Figure 13).

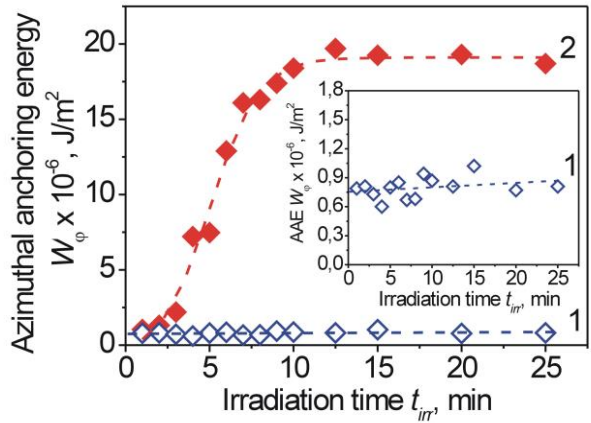

Fig. 13. Dependence of the calculated values of the AAE $W_\varphi$ on irradiation time $t_{irr}$ for the combined twist LC cells in: (1) non-irradiated area of the HAL (opened blue diamond) and (2) irradiated area of the HAL by polarized UV irradiation (solid red diamond). The polarization of UV light $E_\parallel$ is perpendicular to the direction of the nano-grooves of 1D-LSFL $\Omega$. The inset depicts the $W_\varphi(t_{irr})$ in non-irradiated area of the HAL in large scale. Each point of the dependence corresponds to each single LC cell. The average thickness of all LC cells was about 14.8 µm. The intensity of the UV lamp was 1.8 W/m$^2$. The distance between the UV lamp and the tested substrate with HAL was 50 cm. The dashed lines and curve are a guide to the eye.

By comparing the values of the AAE $W_\varphi$ for the HAL being irradiated by UV light and characterized by different polarizations with respect to $\Omega$ (*i.e.* Figure 10 (b) and Figure 13), we can conclude that the use of polarized UV light with $E_\parallel$ perpendicular to the $\Omega$ results in a substantial increase of the AAE $W_\varphi$. In addition, several order of magnitudes of the AAE $W_\varphi$ of HAL irradiated for both polarizations of UV light (*i.e.* for $E_\perp \parallel \Omega$, $W_\varphi$ is within the range of $(0.02 – 0.7) \times 10^{-6}$ J/m² and for $E_\parallel \perp \Omega$, $W_\varphi$ is within the range of $(1 – 19.7) \times 10^{-6}$ J/m²) in comparison with non-irradiated area of the HAL (*i.e.* $W_\varphi$ is about $(0.76 – 0.79) \times 10^{-6}$ J/m²) are observed. In this case, the value of the AAE $W_\varphi$ of HAL, irradiated for 12 min by polarized UV irradiation, when the polarization $E_\parallel$ is perpendicular to the $\Omega$, is about three orders of magnitude higher than the value for the HAL during UV irradiation with the polarization $E_\perp$ parallel to the $\Omega$.

Figure 14 shows a tested sapphire substrate with HAL of the combined twist LC cell disassembled and further wiped clean from LC layer by ethanol. In these photographs, HAL is shown before (a) and after (b) being irradiated for 15 min by polarized UV light $E_\parallel$ perpendicular to the $\Omega$.

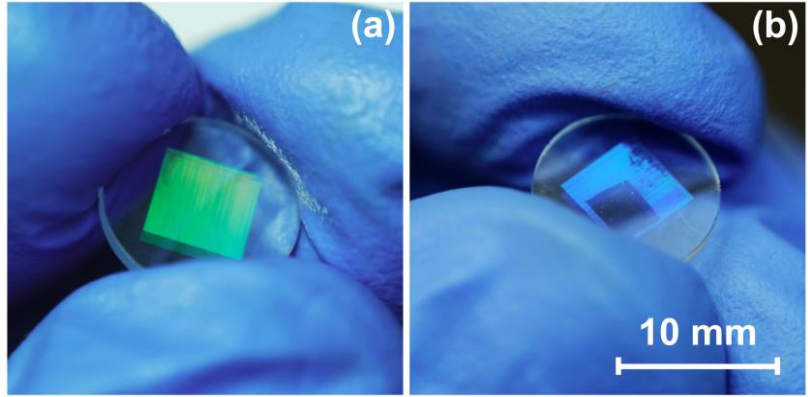

Fig. 14. Photographs of a tested HAL sapphire substrate disassembled of the combined twist LC cell and further wiped clean from LC by ethanol. (a) Before UV irradiation of the HAL, characterized by diffraction scattering of light area of 5 × 5 mm². (b) After irradiation of an area of 3 × 3 mm² of the HAL for 15 min by polarized UV light with polarization $E_\parallel$ perpendicular to the direction of the nano-grooves of 1D-LSFL $\Omega$.

In comparison with the non-irradiated HAL (Figure 14 (a)), being characterized by diffraction scattering of light, due to availability of the nano-grooves obtained by LIPSS processing of the sapphire substrate, the irradiated HAL area of 3 × 3 mm² possesses a high transmittance (Figure 14 (b)). It should be noted that the area with high transmittance was observed for HALs irradiated by polarized UV light, when relative orientation between the polarization $E_\perp$ (or $E_\parallel$) and the $\Omega$ was parallel (or perpendicular). Obviously, the main reason for the absence of the diffraction scattering of the light in the area of 3 × 3 mm² for the UV irradiated HAL is stronger anchoring between LC molecules and the aligning surface in comparison with non-irradiated surface. Considering the different values of the AAE $W_\varphi$ for irradiated and non-irradiated areas of the HAL and also their different degrees of scattering of the light, we can assume that each area of surface will be characterized by different contact angles *β*.

*3.2.5 Contact angle β of the nematic E7 droplet placed on the HAL*

In this subsection we describe the behaviour of the contact angle *β* of the nematic E7 droplet placed on the surface of the HAL characterized by various conditions of the UV irradiation (*i.e.* irradiation time $t_{irr}$, relative orientation between the polarization $E_\perp$ (or $E_\parallel$) and direction of the nano-grooves of 1D-LSFL $\Omega$).

As shown in Ref. [44], the contact angle *β* of the nematic 5CB droplet placed on the irradiated PVCN-F surface by linearly polarized UV light monotonically changes with the exposure time $t_{irr}$. It was also found that the increase of the contact angle *β* occurs during a prolonged exposure time $t_{irr}$ by a powerful UV lamp. In addition, contact angles *β* of droplets of isotropic liquid and nematic E7 placed on various surfaces (graphene, titanium) structured by LIPSS were studied [42,54]. The correlation between the contact angle *β* of nematic E7 droplet placed on the titanium surface processed by LIPSS as well as the azimuthal (AAE) and polar (PAE) anchoring energies of the nanostructured surfaces, characterized by various depths A of nano-grooves, were found [42].

Let us consider the change of the contact angle *β* of the nematic E7 droplet placed on the HAL irradiated by the polarized UV light, when the polarization $E_\perp$ (or $E_\parallel$) is parallel (or perpendicular) to the $\Omega$.

Figure 15 (a) shows the dependence of the contact angle *β* of nematic E7 droplets, placed on the unstructured (solid black squares, line 1) and structured area of the non-irradiated HAL (opened red triangles, curve 2), on the width of the unstructured gap *L*. For the unstructured area (solid black squares, line 1) the value of the contact angle *β* is about 22.4 degrees. However, the availability of nano-grooves leads to an increasing wettability of the surface and as a consequence to the decrease of the value of the contact angle *β* (opened red triangles, curve 2). Obviously, the increase of the width of the unstructured gap *L* leads to an increased contact angle *β* (curve 2). By taking into account the fact that the value of the AAE $W_\varphi$ decreases with the increasing width of the unstructured gap *L* [25], we can conclude that for the HAL a similar correlation between the contact angle *β* and the AAE $W_\varphi$ for a *zebra*-like surface of the sapphire substrate having a certain width of the unstructured gap *L* exists, as previously found in Ref. [42] for Ti surface processed by LIPSS.

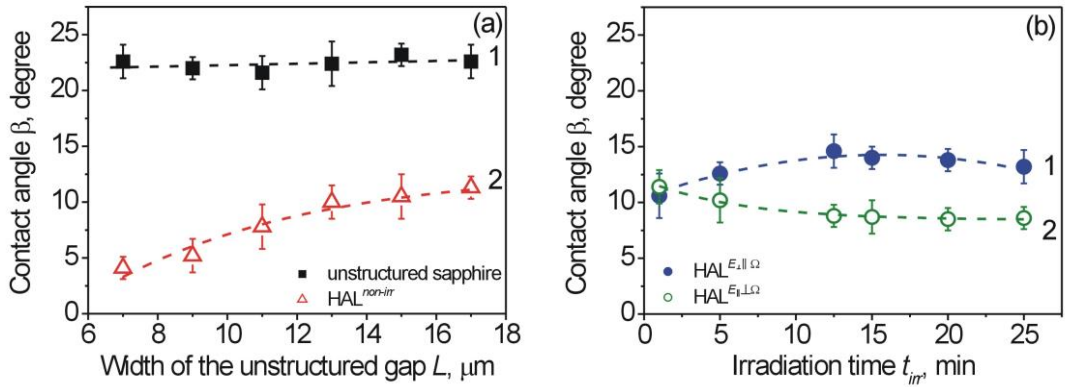

Fig. 15. (a) The dependence of contact angle *β* of nematic E7 droplets placed on the unstructured (solid black squares, line 1) and structured area of the non-irradiated HAL (opened red triangles, curve 2) on the width of the unstructured gap *L*. (b) Irradiation time dependence of contact angle *β* of nematic E7 droplets placed on the HAL area, irradiated by polarized UV light with the polarization $E_\perp$ parallel (solid blue circles, curve 1) and $E_\parallel$ perpendicular (opened green circles, curve 2) to the direction of the nano-grooves of 1D-LSFL $\Omega$ HAL characterized by the 17 µm width of the unstructured gap. The intensity of the UV lamp was 1.8 W/m$^2$. The distance between the UV lamp and the sapphire substrate with HAL was 50 cm. The dashed curves and lines are a guide to the eye.

Figure 15 (b) shows the change of the contact angle $\beta$ for different conditions of the irradiation of the HAL, as schematically illustrated in Figure 1 (b)-(e). Curve 1 (solid blue circles) depicts the irradiation time dependence of the contact angle $\beta$ at the surface of the HAL irradiated by polarized UV light $E_\perp$ being parallel to the $\Omega$. The contact angle $\beta$ of the nematic droplet increases for the surfaces of the HAL for irradiation times shorter then 12 min. The dependence $\beta(t_{irr})$ shows that the long-term exposure of the HAL (solid blue circles, curve 1) leads to a decrease of the contact angle $\beta$ of the nematic E7 droplet. By taking into account the dependence of the twist angle $\varphi$ on irradiation time $t_{irr}$ (Figure 10 (b)), the same correlation of the contact angle $\beta$ is observed. For instance, the nematic E7 droplet placed on the surface of the HAL, possessing weak anchoring energy, has the maximum value of contact angle $\beta$ and vice versa. The correlation between AAE $W_\varphi$ and contact angle $\beta$ is shown in Figure 16 by line 1.

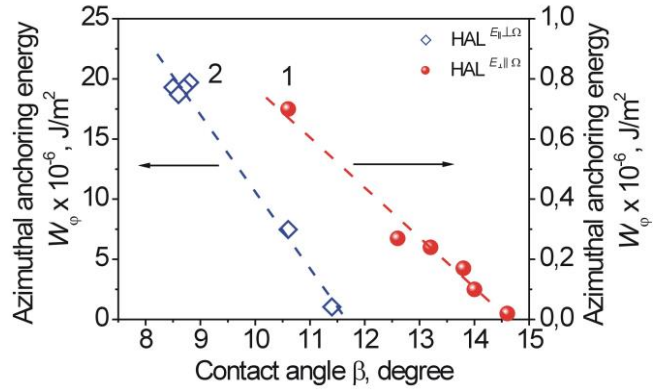

Fig. 16. Correlation between AAE $W_\varphi$ and contact angle $\beta$ for the HAL, irradiated by UV light with the polarization: (1) $E_\perp$ parallel (solid blue circles) and (2) $E_\parallel$ perpendicular (opened green circles) to the direction of the nano-grooves of 1D-LSFL $\Omega$. The intensity of the UV lamp was 1.8 W/m$^2$. The distance between the UV lamp and the sapphire substrate with HAL was 50 cm.

For the HAL irradiated by UV light with the polarization $E_\parallel$ perpendicular to the $\Omega$, a monotonically decrease of the contact angle $\beta$ with irradiation time $t_{irr}$ is observed (opened green circles, Figure 16 (b)). As shown in Figure 13, the monotonically growth of the AAE $W_\varphi$ of irradiated HAL is observed for short exposure time ($t_{irr} < 12$ min). It should be noted that during prolonged UV exposure ($t_{irr} > 12$ min) both, the AAE $W_\varphi$ (Figure 13) and the contact angle $\beta$ (curve 2, Figure 15 (b)) has remained nearly unchanged. This is evidenced by set of points (line 2) shown in Figure 16.

### 4 Conclusions

In general, we can conclude that the hybrid aligning layer (HAL) can be created by using any high-quality structured substrate obtained by LIPSS technique covered with any photoaligning polymer layer. These combined aligning layers can be characterized by phototuning of main alignment properties such as the easy axis of orientation, twist angle and anchoring energy.

In this work, we illustrated aligning properties of the hybrid aligning layer (HAL) by example of a *zebra*-like surface of the sapphire substrate covered by photosensitive polymer PVCN-F. It was experimentally shown, that in aligning nematic LCs the usage of the HAL could be very useful for different applications. Such surfaces can provide both the alignment caused by the availability of the nano-grooves formed by LIPSS and the light-controllable alignment owing to the use of the photosensitive PVCN-F layer.

We have also shown that the use of different methods and conditions during treatment of the HAL, namely the variation of the width $L$ (unstructured gap of a *zebra*-like surface) and the irradiation of the HAL by polarized UV light in relation to the direction of the orientation of the nano-grooves of 1D-LSFL; leads to photoinduced changes of the value of AAE $W_\varphi$ and twist angle $\varphi$ of combined twist LC cell.

Furthermore, it was experimentally shown that AAE $W_\varphi$ of the HAL can be weakened when the irradiation of the HAL occurs by UV light with polarization $E_\perp$ parallel to the direction of the nano-grooves of 1D-LSFL $\Omega$. We also studied the influence of the width of the unstructured gap $L$ of a *zebra*-like surface of the HAL on the value of the AAE $W_\varphi$ of aligning layer. For the irradiated HAL, characterized by a small width of the unstructured gap $L = 7$ µm, a decreasing AAE $W_\varphi$ in the range from 1.24 to 0.18 $\times 10^{-6}$ J/m$^2$ is observed. In case of the unstructured gap $L = 13$ and 17 µm, both a decreasing value of the AAE $W_\varphi$ of HAL during short exposure time ($t_{irr} < 12$ min) and an increasing value of the AAE $W_\varphi$ of HAL during longer exposures ($t_{irr} > 12$ min) were found. The use of the HAL, characterized by unstructured gap $L = 13$ and 17 µm, irradiated by polarized UV light for ~ 12 min, leads to the appearance of the surface with very weak anchoring, possessing by near-zero value of the AAE $W_\varphi$ ~ 2 $\times 10^{-8}$ J/m$^2$. We believe that the decreasing/increasing of the AAE $W_\varphi$ (or twist angle $\varphi$) of the HAL occurs due to the competition between two alignments, namely, on the one hand, caused by the availability of the nano-grooves of the sapphire surface and, on the other hand, due to the photosensitive PVCN-F layer, that covers of the nano-grooves of sapphire surface.

It was further found that during irradiation of the HAL by polarized UV light, when the polarization $E_\parallel$ is perpendicular to the direction of the nano-grooves of 1D-LSFL $\Omega$, a significant growth of the AAE $W_\varphi$ occurs (*e.g.* about 19.7 $\times 10^{-6}$ J/m$^2$ for unstructured gap $L = 17$ µm). The main reason for this is the fact that the direction of alignment caused by the nano-grooves of a *zebra*-like surface of sapphire surface of the HAL coincides with the direction of the photoinduced alignment due to UV irradiated PVCN-F layer.

During UV irradiation of the HAL, the value of the AAE $W_\varphi$ can be controlled in the wide range from 2 $\times 10^{-8}$ to 19.7 $\times 10^{-6}$ J/m$^2$ by changing the polarization of UV light in relation to the direction of the nano-grooves of 1D-LSFL $\Omega$ and exposure time.

**Acknowledgments**

The authors thank W. Becker (Merck, Darmstadt, Germany) for his providing of nematic liquid crystal E7, A. Kratz (Merck, Germany) for her produce of the Licristal brochure, Prof. I. Gerus (V. P. Kukhar Institute of Bioorganic Chemistry and Petrochemistry of the NAS of Ukraine) for the kind provision of the PVCN-F photopolymer, Dr. A. Buchnev (Cambridge, England) who had provided us with polymer PI2555 and field service specialist IV V. Danylyuk (Dish LLC, USA) for his gift of some Laboratory equipments. The authors I.G., D.B., Z.K., S.M. and P.L. thank warriors of the Armed Forces of Ukraine who are protecting us during full-scale war in Ukraine. The authors thank Prof. L. Lisetski (Institute for Scintillation Materials of STC "Institute for Single Crystals" of the NAS of Ukraine), Dr. S. Lukyanets and Dr. S. Chernyshuk (Institute of Physics of the NAS of Ukraine) for the helpful discussions. In addition, the authors are incredibly grateful to the Referees for the constructive comments.

**Figure captions**

Fig. 1. Schematic of the studied substrates and their irradiation. (a) HAL, consisting of sapphire substrate (blue colour) coated by photosensitive PVCN-F layer (orange colour) before UV irradiation. The direction of the LSFL ($\Omega$) is shown by red and black arrow. $\vec{n}_{HAL}$ (black and white arrow) is the orientation of the easy axis of the HAL. HAL irradiated by polarized UV light with the polarization $E_\perp$ being parallel to the nano-grooves with their period $\Lambda$ and depth A over time $t_{irr}^1$ (b) and $t_{irr}^2 > t_{irr}^1$ (c). HAL irradiated by polarized UV light with the polarization $E_\parallel$ being perpendicular to the nano-grooves over time $t_{irr}^1$ (d) and $t_{irr}^2 > t_{irr}^1$ (e). $\vec{n}_{1D-LSFL}$ (black and yellow arrow) and $\vec{n}_{PVCN-F}$ (black and blue arrow) are directions of easy axis caused by structured sapphire substrate obtained by LIPSS and PVCN-F layer after UV-irradiation, respectively.

Fig. 2. Schematic setup of irradiation of the hybrid aligning layer based on the LIPSS on the sapphire and photoinduced aligning polymeric layer of the PVCN-F. The partial image of HAL schematically shows a *zebra*-like surface of sapphire substrate structured by LIPSS. A *zebra*-like surface is characterized by LSFL lines with 23 μm width, fixed period of nano-grooves $\Lambda_{LSFL} = 980$ nm and different widths of unstructured gap *L* (7, 13 and 17 μm). The intensity of the UV lamp was 1.8 W/m$^2$ on the substrate. The distance between the UV lamp and the substrate was 50 cm.

Fig. 3. Photographs of the combined twist LC cell, assembled from reference substrate with PI2555 layer, rubbed $N_{rubb} = 15$ at an angle of 45 degree, and tested substrate with PVCN-F coating between: (a) parallel and (b) crossed polarizer P and analyser A. Each area of PVCN-F layer was irradiated by: 1) $t_{irr}^1 = 1$ min; 2) $t_{irr}^2 = 3$ min; 3) $t_{irr}^3 = 6$ min; 4) $t_{irr}^4 = 10$ min; 5) $t_{irr}^5 = 30$ min. The intensity of the UV lamp was 1.8 W/m$^2$. The distance between the UV lamp and the tested substrate with PVCN-F layer was 50 cm. The plane of polarization of Glan-Thompson prism $E_\parallel$ is shown by blue arrow. The plane of polarization of the polarizer P coincides with the rubbing direction (black and green arrow) on the reference substrate. The thickness of the LC cell was 13.3 μm.

Fig. 4. Dependence of the twist angle $\varphi$ of the combined twist LC cell (a) and AAE $W_\varphi$ of the glass substrate covered with PVCN-F film (b) on the irradiation time $t_{irr}$. The intensity of the UV lamp was 1.8 W/m$^2$. The distance between the UV lamp and the tested substrate with PVCN-F layer was 50 cm. The thickness of the LC cell was about 13.3 μm. The dashed curve is a guide to the eye.

Fig. 5. (a) AFM image and (b) cross-section of the 1D-LSFL taken from AFM shown for the HAL, possessing periodic nano-grooves with period $\Lambda \sim 972$ nm and depth A $\sim 50$ nm. The cross-section is normalized to the X-axis. Photographs of combined twist LC cells, consisting of the reference unidirectionally rubbed glass substrate with PI2555 film coating and the tested substrates with: (c) a *zebra*-like surface coated with PVCN-F film, *i.e.,* HAL, and (d) the unstructured sapphire surface covered with PVCN-F before irradiation by UV light. Red and black arrow shows direction of nano-grooves of the LSFL $\Omega$. Black and green arrow shows the rubbing direction of the reference PI2555 film coating glass substrate. Thickness of LC cells was about 20 μm.

Fig. 6. AFM images and the cross-section of the 1D-LSFL taken from AFM reveal for the HAL irradiated by polarized UV-light with: (a), (b) the polarization $E_\perp$ parallel to the direction of the nano-grooves $\Omega$ of 1D-LSFL (period $\Lambda \sim 972$ nm and depth A $\sim 38$ nm) (c), (d) the polarization $E_\parallel$ perpendicular to the direction of the nano-grooves $\Omega$ of 1D-LSFL (period $\Lambda \sim 972$ nm and depth A $\sim 62$ nm). The irradiation time of the HAL was 15 min. The intensity of the UV lamp was 1.8 W/m$^2$. The distance between the UV lamp and the tested substrate with PVCN-F layer was 50 cm. The cross-sections of the areas with 1D-LSFL were normalized to the X-axis.

Fig. 7. The irradiation time dependence of the twist angle $\varphi$ (a) and AAE $W_\varphi$ (b) of the combined twist LC cell, consisting of the tested substrate, having both non-irradiated HAL (opened black symbols, lines 1) and HAL (solid symbols, curves 2), irradiated by polarized UV light with the polarization $E_\perp$ parallel to the $\Omega$. HAL with a *zebra*-like surface of the structured sapphire substrate has 7 μm width of the unstructured gap. Each point corresponds to a single LC cell. The average thickness of all LC cells was about 15.5 μm. The intensity of the UV lamp was 1.8 W/m$^2$. The distance between the UV lamp and the tested substrate with HAL was 50 cm. The dashed lines and curves are a guide to the eye.

Fig. 8. The dependence of the twist angle $\varphi^{PVCN-F}$ of the combined twist LC cell, consisting of the tested PVCN-F layer coated onto a glass substrate (opened blue circles, curve 1) and the difference $\phi$ between the twist angle values of the combined twist LC cell consisting of the tested substrate with HAL having non-irradiated and irradiated areas (solid green triangles, curve 2), on the irradiation time $t_{irr}$. The inset shows $\Delta\varphi(t_{irr})$. The dashed curves are a guide to the eye.

Fig. 9. The irradiation time dependence of the twist angle $\varphi$ of the combined twist LC cells consisting of the tested sapphire substrate with HAL. The twist angle of LC cells for non-irradiated area of the HAL (opened black symbols, lines 1) is: (a) $\varphi \sim 31.5°$ for $L = 13$ μm and (b) $\varphi \sim 27.5°$ for $L = 17$ μm. The twist angle of the LC cells with the HAL irradiated with polarized UV light when the polarization $E_\perp$ is parallel to the $\Omega$ (spheres, curves 2). Solid red stars correspond to twist LC cells having substrate with HAL irradiated by UV light for 12 min. Each point within the plots corresponds to each single LC cell. The average thickness of all LC cells was about 15.4 μm. The intensity of the UV lamp was 1.8 W/m$^2$. The distance between the UV lamp and the tested substrate with HAL was 50 cm. The dashed lines and curves are a guide to the eye.

Fig. 10. Dependence of the AAE $W_\varphi$ of the combined twist LC cells on irradiation time $t_{irr}$. LC cells consist of the tested substrate with HAL, possessing structured sapphire surface with the width of the unstructured gap: (a) $L = 13$ μm and (b) $L = 17$ μm. The AAE $W_\varphi$ of the non-irradiated HAL (opened black squares, lines 1) is: (a) $\sim 1.1 \times 10^{-6}$ J/m$^2$ for $L = 13$ μm and (b) $\sim 0.76 \times 10^{-6}$ J/m$^2$ for $L = 17$ μm. The minimum of the value of AAE $W_\varphi$ for the HAL irradiated within 12 min by UV light with the polarization $E_\perp$ parallel to the direction of LSFL $\Omega$ (solid triangles, curves 2) is: (a) $\sim 3.2 \times 10^{-8}$ J/m$^2$ for $L = 13$ μm and (b) $\sim 2 \times 10^{-8}$ J/m$^2$ for $L = 17$ μm. Each point of the dependence corresponds to each single LC cell. The average thickness of all LC cells was about 15.4 μm. The dashed lines and curves are a guide to the eye.

Fig. 11. Photograph of the combined twist LC cell, consisting of the reference substrate with rubbed PI2555 layer ($N_{rubb}$ = 15) and tested sapphire substrate with HAL, possessing the 17 µm width of unstructured gap, placed between crossed polarizer (P) and analyzer (A) at angles: (a) 27.5 and (b) 90 degrees. HAL possesses two areas: (1) non-irradiated and (2) irradiated by linearly polarized UV light (*i.e.* when the polarization $E_\perp$ is parallel to the $\Omega$) for 12 min. The thickness of the LC cell was 15.5 µm.

Fig. 12. Dependence of the twist angles $\varphi$ on the irradiation time $t_{irr}$ for the combined twist LC cells consisting of the reference substrate (rubbed PI2555 layer) and the tested sapphire substrate with: (a) irradiated by UV light with the polarization $E_\parallel$ perpendicular to the $\Omega$ and (b) non-irradiated areas of the HAL. The change of twist angle for: 23 µm width nano-grooves 1D-LSFL (solid blue symbols, curves 1), 17 µm width unstructured gap (opened green symbols, curves 2) and their average value (spheres symbols, curves 3). HAL consists of a *zebra*-like surface of the sapphire substrate (*i.e.* 23 µm 1D-LSFL and unstructured gap $L$ = 17 µm) covered with the PVCN-F layer. Each point of the dependence corresponds to each single LC cell. The average thickness of all LC cells was about 14.8 µm. The intensity of the UV lamp was 1.8 W/m$^2$. The distance between the UV lamp and the tested substrate with HAL was 50 cm. The dashed curves and lines and curves are a guide to the eye.

Fig. 13. Dependence of the calculated values of the AAE $W_\varphi$ on irradiation time $t_{irr}$ for the combined twist LC cells in: (1) non-irradiated area of the HAL (opened blue diamond) and (2) irradiated area of the HAL by polarized UV irradiation (solid red diamond). The polarization of UV light $E_\parallel$ is perpendicular to the direction of the nano-grooves of 1D-LSFL $\Omega$. The inset depicts the $W_\varphi(t_{irr})$ in non-irradiated area of the HAL in large scale. Each point of the dependence corresponds to each single LC cell. The average thickness of all LC cells was about 14.8 µm. The intensity of the UV lamp was 1.8 W/m$^2$. The distance between the UV lamp and the tested substrate with HAL was 50 cm. The dashed lines and curve are a guide to the eye.

Fig. 14. Photographs of a tested HAL sapphire substrate disassembled of the combined twist LC cell and further wiped clean from LC by ethanol. (a) Before UV irradiation of the HAL, characterized by diffraction scattering of light area of 5 × 5 mm$^2$. (b) After irradiation of an area of 3 × 3 mm$^2$ of the HAL for 15 min by polarized UV light with polarization $E_\parallel$ perpendicular to the direction of the nano-grooves of 1D-LSFL $\Omega$.

Fig. 15. (a) The dependence of contact angle $\beta$ of nematic E7 droplets placed on the unstructured (solid black squares, line 1) and structured area of the non-irradiated HAL (opened red triangles, curve 2) on the width of the unstructured gap $L$. (b) Irradiation time dependence of contact angle $\beta$ of nematic E7 droplets placed on the HAL area, irradiated by polarized UV light with the polarization $E_\perp$ parallel (solid blue circles, curve 1) and $E_\parallel$ perpendicular (opened green circles, curve 2) to the direction of the nano-grooves of 1D-LSFL $\Omega$ HAL characterized by the 17 µm width of the unstructured gap. The intensity of the UV lamp was 1.8 W/m$^2$. The distance between the UV lamp and the sapphire substrate with HAL was 50 cm. The dashed curves and lines are a guide to the eye.

Fig. 16. Correlation between AAE $W_\varphi$ and contact angle $\beta$ for the HAL, irradiated by UV light with the polarization: (1) $E_\perp$ parallel (solid blue circles) and (2) $E_\parallel$ perpendicular (opened green circles) to the direction of the nano-grooves of 1D-LSFL $\Omega$. The intensity of the UV lamp was 1.8 W/m$^2$. The distance between the UV lamp and the sapphire substrate with HAL was 50 cm.